\documentstyle[twocolumn,prd,aps]{revtex}

\evensidemargin=\oddsidemargin
\begin{document}

\title{Testing Technicolor Models via Top Quark Pair Production in High
Energy Photon Collisions}
\author{ Hong-Yi Zhou$^{a}$~~Yu-Ping Kuang$^{a}$~~Chong-Xing Yue$^{b}$,
Hua Wang$^{a}$,~~ Gong-Ru Lu $^{b}$}
\address{
$^a$ Department of Physics,Tsinghua University, Beijing, 
 100084, P. R. of China.\\
$^b$  Physics Department, Henan Normal University, Xin Xiang, 
 Henan 453002,  P. R. of China 
}

\maketitle
\vspace{0.8cm}

\abstract{
~~~ Pseudo-Goldstone boson contributions to $t\bar{t}~$ 
production rates in technicolor models with and without
topcolor at the $\sqrt{s}=0.5~{\rm and}~1.5$~TeV photon colliders 
and hadron colliders are reviewed.   
For reasonable ranges of the parameters, 
the contributions  are large enough to be experimentally observable. Models 
with topcolor, without topcolor and the MSSM with $\tan\beta\geq 1$ can be 
experimentally distinguished. 
}

\pacs{}





\begin{center}
{\bf I. Introduction}
\end{center}

Technicolor (TC) theory is  an 
attractive idea of dynamical 
Electroweak Symmetry Breaking(EWSB) which avoids the shortcomings arising 
from elementary scalar fields. There are  
extended technicolor (ETC) theory to give quark and 
lepton masses;walking technicolor (WTC) or multiscale 
walking technicolor theory(MTC) to give small FCNC; 
topcolor-assisted technicolor (TOPCTC) theory to account 
for large top quark mass.
These theories all contain certain pseudo Goldstone bosons (PGB's) 
in the few hundred GeV region. Testing their 
virtual effects are complimentary to the direct searches for them. 

The top quark couples strongly to the PGB's due to its large mass, so that the 
virtual effects  would be more apparent 
in processes with the top quark than with any other light quarks. 
Top quark pair can be produced at various high energy colliders.
Of special interest is  the suggested future
TeV energy photon colliders which can be realized by laser back 
scattering technique from $e^+e^-$ colliders. 

The PGB's one-loop virtual effects of 
some typical TC models in the top quark pair production via 
$\gamma\gamma\to t\bar t$ have been studied in Ref.\cite{ZHY} in which
the Appelquist-Terning
one-family model(AT) \cite{AT} was taken as a typical example of the  TC 
models without assisted by topcolor;
the original TOPCTC model by Hill \cite{TOPC} and the  top-color 
assisted multiscale TC model (TOPCMTC)
\cite{Lane} were taken as typical models assisted by topcolor. 
For the TOPCMTC, the  resonance contributions of the 
neutral PGB's to $t\bar t$ production at hadron colliders have been 
studied in Ref.\cite{ZHY1}.

Here we give a brief review of these studies.

\vspace{0.5cm}
\begin{center}
{\bf II. PGB corrections to $~t\bar{t}$ production  at photon colliders 
in the one family  model}
\end{center}

We consider the Appelquist-Terning one-family model(AT) \cite{AT}.
In this model, The TC group  is taken to be 
$~SU(2)_{TC}~$ which minimizes the $~S~$ parameter. There are 36 TC PGB's
composed of weak $~SU(2)_W~$ doublets of techniquarks Q and technileptons L. 
The relevant PGB's in this study are the color-octet $\Pi_a^0~[SU(2)_W$-
singlet] and $\Pi_a^\alpha~[SU(2)_W$-triplet]  composed of the techniquarks 
$~Q~$(the color-singlet PGB's in this
model are mainly composed of technileptons $L$, so that they are
irrelevant to the $t\bar{t}$ production). 
The decay constants of these  PGB's is $~F_Q =140~$GeV . The
masses of these PGB's are model dependent \cite{AT}. 
The mass of $\Pi^\alpha_a$ is taken to be in the range
$~250~{\rm GeV}<m^\alpha_a<500~{\rm GeV}~$. We also take the mass
of $\Pi^0_a$ in the same range in our calculation.



The total cross section $~\sigma(s)~$ of the production of $~t\bar{t}~$ in 
$~\gamma\gamma~$ collisions is obtained by folding  the elementary cross 
section $~\sigma(\hat{s})~$ for the subprocess $~\gamma \gamma \to t \bar{t}~$ 
with the photon luminosity at the $ e^+ e^- $ collider, i.e.
\begin{eqnarray}                                
& &\sigma(s)=\int^{x_{max}}_{2m_t/\sqrt{\hat{s}}}dz \frac{dL_{\gamma\gamma}} 
{d z}\sigma(\hat{s)} \,,
\end{eqnarray}
where $~\sqrt{s}(\sqrt{\hat{s}})~$ is the $~e^+e^- (\gamma \gamma)~$ center-of
-mass energy and $~dL_{\gamma \gamma}/dz~$ is the photon luminosity defined as
\begin{eqnarray}                                
\frac{dL_{\gamma\gamma}}{d z} = 2z \int^{x_{max}}_{z^2 /x_{max}}
 \frac{dx}{x}F_{\gamma/e}(x) F_{\gamma/e}(z^2/x)\,.
\end{eqnarray}
The following  cuts on the rapidity $~y~$ and the transverse 
momentum $~p_T~$ are taken in all the calculations: 
\begin{eqnarray}                            
|y|< 2.5,~~~~~~~~~~p_T> 20~\rm{GeV}\,.
\end{eqnarray}

We take $m_t=176$~GeV, $m_b=4.9$~GeV, and we take
$\alpha_{em}(m_Z^2)=1/128.8$ with the one-loop running formula to determine the 
electromagnetic fine structure constant $\alpha_{em}$ at the desired scale.
The result of the tree-level cross sections are 

$\sigma_0=57.77$~fb for $\sqrt{s}=0.5$ TeV;
  
$\sigma_0=535.4$~fb for $\sqrt{s}=1.5$ TeV.
 
To see the main feature of the TC PGB contributions to the cross section, we
simply take $m_{\Pi^0_a}=m_{\Pi^3_a}=m_{\Pi^{\pm}_a}\equiv m_{\Pi_a}$ to
calculate the correction $\Delta \sigma$. We find that for 
$\sqrt{s}=0.5$ TeV, the relative correction $\Delta\sigma/\sigma_0$ 
is of the order of $10\%$
which is about one order of magnitude 
larger than that in the MSSM with $\tan\beta\geq 1$
(which is about one percent\cite{MSSM}). For 
$\sqrt{s}=1.5$~TeV, the relative corrections are around $(4-10)\%$ which is 
also larger than that in the MSSM with $\tan\beta\geq 1$.

For estimating the event rates, we  take an integrated luminosity of 
\begin{eqnarray}                             
\int {\cal L}dt=50~{\rm fb}^{-1},~~~~ {\rm for}~\sqrt{s}=0.5~{\rm TeV}\,,\nonumber\\
\int {\cal L}dt=100~{\rm fb}^{-1},~~~~ {\rm for}~\sqrt{s}=1.5~{\rm TeV}\,,
\end{eqnarray} 
which corresponds to a one year run at the NLC. There will be about 
$2500$ events for $\sqrt{s}=0.5$~TeV and $25000$ events for
$\sqrt{s}=1.5$~TeV. The statistical uncertainty at $95\%$ C.L. is then
around $4\%$ for $\sqrt{s}=0.5$~TeV and $1.2\%$ for $\sqrt{s}=1.5$~TeV. 
Therefore {\it this model can be experimentally distinguished from the
MSSM model with $\tan\beta\geq 1$ in $\gamma\gamma\to t\bar{t}$ at the future 
photon collider}.

\vspace{0.5cm}
\begin{center}
 {\bf III. $t\bar{t}$ Production cross sections in TOPCTC Models }
\end{center}
\vspace{0.4cm}
\null\noindent
{\bf 1.}~The Original TOPCTC Model \cite{TOPC}

For TOPCTC models, we first consider the original TOPCTC model proposed
by Hill \cite{TOPC}. In this model, there are 60 TC PGB's in the TC sector
with the decay constant $f_{\Pi}=120~$GeV and three top-pions $\Pi_t^{0}$
, $\Pi_t^{\pm}$ in the topcolor sector with the decay constant
$f_{\Pi_t}=50~$GeV \cite{TOPC}. The top quark mass $m_t$ is mainly provided 
by the topcolor sector, while the TC sector only provide a small portion of it,
say $m'_t\sim 5-24$~GeV \cite{TOPC}. The mass of the top-pion
depends on a parameter in the model \cite{TOPC}. For reasonable
values of the parameter, $m_{\Pi_t}$ is around $200~$GeV \cite{TOPC}.
The prediction of the light top-pion is the characteristic feature of
the TOPCTC models and this is the main difference between this kind of models 
and TC models without topcolor. In the following calculation, we 
would rather take a slightly larger range, $~180~{\rm GeV}\leq m_{\Pi_t}
\leq 300~$GeV, to see its effect, and we shall take the masses of the color-
singlet TC PGB's to vary in the range $100-325~$GeV.

We find that the cross section correction in photon colliders 
are considerably larger than those  in the AT model. 
This is because of the large top pion contributions.
Taking the integrated luminosity in the above, we have around $1000$ events
for $\sqrt{s}=0.5$~TeV and around $40000$ events for $\sqrt{s}=1.5$~TeV.
The corresponding statistical uncertainties at the $95\%$ C.L. are then 
$6\%$ and $1\%$, respectively. We get the conclusion 
that  {\it this model is experimentally 
distinguishable from the Appelquist-Terning model and the MSSM with 
$\tan\beta\geq 1$ in $\gamma\gamma\to t\bar{t}$}.\\

\null\noindent
{\bf 2.}~The TOPCMTC Model \cite{Lane}

In this model, the technicolor sector is multiscale. 
It is different from the original TOPCTC model 
mainly by the change of the TC sector in which $f_{\Pi}=40~$GeV instead of
the original $f_{\Pi}=120~$GeV.

For $\sqrt{s}=0.5$ TeV, the results of this
model at photon colliders are close to those in the original TOPCTC model. 
For $\sqrt{s}=1.5$ TeV, the corrections are
much larger than those in the original TOPCTC model, 
especially with large $m_t'$. From the results 
we conclude that  {\it even the difference between the original 
TOPCTC model and the TOPCMTC model can be clearly observed in the 
$\gamma\gamma\to t\bar{t}$ experiment at the $\sqrt{s}=1.5$~TeV photon 
collider}.

The  PGB corrections to $t\bar t$ production 
at the Fermilab Tevatron and the CERN LHC in this model are also 
studied\cite{ZHY1}(for other models,see Ref.\cite{Lane1}). 
We find that the Tevatron experimental results can 
give constraits on the parameters. At the LHC, the resonance peaks 
in the production are significant. Even though we can obtain 
much more better statistical errors at the LHC, the systematic 
error of the cross section is about $5\%$. Whether it is possible 
to distinguish the different TC models at the LHC is under investigation.

\begin{center}
{\bf Acknowledgements} 
\end{center}

H.Y.Zhou is financially supported by the Alexander von Humboldt  
Foundation of Germany.




\vspace{0.5cm}
\begin{center}
{\bf Reference}
\end{center}
\begin{enumerate}

\bibitem{ZHY}
Hong-Yi Zhou,~~Yu-Ping Kuang,~~Chong-Xing Yue,  
Hua Wang,~~ Gong-Ru Lu, Phys. Rev. {\bf D 57} (1998) 4205. 


\bibitem {AT}
T. Appelquist and J. Terning, Phys. Lett. {\bf B315}
(1993) 139 .


\bibitem {TOPC}
C. T. Hill, Phys. Lett. {\bf B345} (1995) 483; K. Lane and E. Eichten, 
Phys. Lett. {\bf B352} (1995) 382; G. Buchalla, G.Burdman, C. T. Hill and 
D. Kominis, Phys. Rev. D{\bf 53} (1996) 5185, 
FERMILAB-PUB-95/322-T.


\bibitem{Lane}
K. Lane, Phys. Lett. {\bf B357}, (1995) 624. 

\bibitem{ZHY1}
Chong-Xing Yue, Hong-Yi Zhou, ~~Yu-Ping Kuang, ~~ Gong-Ru Lu,  
Phys. Rev. {\bf D 55} (1997) 5541. 



\bibitem {MSSM}
C.-S. Li, J.-M. Yang, Y.-L. Zhu, and H.-Y. Zhou, Phys. Rev. D{\bf 54}  
(1996) 4662; H. Wang, C.-S. Li, H.-Y. Zhou, and Y.-P. Kuang, Phys. Rev. 
D{\bf 54} (1996) 4374.

\bibitem{Lane1}
E. Eichten and K. Lane, Phys. Lett. B{\bf 327} (1994) 129;
G.L. Lu, Y.Hua, J.M. Yang, and X.L. Wang, Phys. Rev. D {\bf 54} (1996) 
1083. 

\end{enumerate}
\end{document}